\documentclass[nonacm,acmsmall,screen]{acmart}

\acmConference[PL'18]{ACM SIGPLAN Conference on Programming Languages}{January 01--03, 2018}{New York, NY, USA}
\acmYear{2018}
\acmISBN{} 
\acmDOI{} 
\startPage{1}

\setcopyright{none}

\usepackage{physics}
\usepackage{listings}
\usepackage{idrislang}
\usepackage{tikz}
\usepackage{hyperref}
\newcommand{\name}{Qimaera}%

\newcommand{\secref}[1]{\S\ref{#1}}

\newcommand{\defeq}{\stackrel{\textrm{{\scriptsize def}}}{=}}
\newcommand{\eqdef}{\defeq}

\theoremstyle{plain}
\newtheorem{theorem}{Theorem}[subsection]

\theoremstyle{definition}

\newtheorem{example}[theorem]{Example}

\theoremstyle{remark}
\newtheorem{remark}[theorem]{Remark}

\usetikzlibrary{decorations.pathmorphing}
\usetikzlibrary{decorations.markings}
\usetikzlibrary{decorations.pathreplacing}
\usetikzlibrary{arrows}
\usetikzlibrary{shapes}

\pgfdeclarelayer{edgelayer}
\pgfdeclarelayer{nodelayer}
\pgfsetlayers{edgelayer,nodelayer,main}

\tikzset{
cross/.style={path picture={
\draw[-,black](path picture bounding box.north) -- (path picture bounding box.south) (path picture bounding box.west) -- (path picture bounding box.east);
}}}

\tikzstyle{braceedge}=[decorate,decoration={brace,amplitude=10pt}]
\tikzstyle{square box}=[rectangle,fill=white,draw=black,minimum height=6mm,minimum width=6mm,yshift=0.7mm]
\tikzstyle{box}=[rectangle,fill=white,draw=black]
\tikzstyle{dot}=[circle,fill=black,draw=black,inner sep=1.5pt]
\tikzstyle{target}=[{draw,circle,cross,minimum width=0.3 cm}]

\tikzstyle{none}=[inner sep=0pt]
\tikzstyle{empty}=[rectangle,fill=none,draw=none]
\tikzstyle{scaled}=[rectangle,fill=none,draw=none, font=\small]

\tikzstyle{to}=[->,draw=black]
\tikzstyle{naturalto}=[-{Implies},double distance=1.5pt]
\tikzstyle{hook}=[right hook->, draw=black]
\tikzstyle{blueArr}=[->, draw=blue]

\tikzstyle{equal-arrow}=[double equal sign distance]

\tikzstyle{every picture}=[baseline=-0.25em]

\newcommand{\stikz}[2][1]{\scalebox{#1}{
\InputIfFileExists{#2}{}{\input{./tikz/#2}}
}}
\newcommand{\cstikz}[2][1]{\begin{center}\stikz[#1]{#2}\end{center}}

\begin{document}

\title{Qimaera: Type-safe (Variational) Quantum Programming in Idris}

\author{Liliane-Joy Dandy}
\affiliation{
  \institution{Ecole Polytechnique/EPFL}
  \country{France/Switzerland}
}
\author{Emmanuel Jeandel}
\affiliation{%
  \institution{Université de Lorraine, LORIA}
  \country{France}
}
\author{Vladimir Zamdzhiev}
\affiliation{%
  \institution{Inria}
  \country{France}
}


\begin{abstract}
  Variational Quantum Algorithms are hybrid classical-quantum algorithms where
  classical and quantum computation work in tandem to solve computational
  problems. These algorithms create interesting challenges for the design of
  suitable programming languages. In this paper we introduce Qimaera, which is
  a set of libraries for the Idris 2 programming language that enable the
  programmer to implement (variational) quantum algorithms where the full power
  of the elegant Idris language works in synchrony with quantum programming
  primitives that we introduce. The two key ingredients of Idris that make this
  possible are (1) dependent types which allow us to implement unitary (i.e.
  reversible and controllable) quantum operations; and (2) linearity which
  allows us to enforce fine-grained control over the execution of quantum
  operations that ensures compliance with the laws of quantum mechanics. We
  demonstrate that Qimaera is suitable for variational quantum programming by
  providing implementations of the two most prominent variational quantum
  algorithms -- QAOA and VQE. To the best of our knowledge, this is the first
  implementation of these algorithms that has been achieved in a
  \emph{type-safe} framework.
  
\end{abstract}


\keywords{variational quantum programming, dependent types, linear types}


\maketitle

\section{Introduction}
\label{sec:intro}

\emph{Variational Quantum Algorithms} \cite{vqa1,vqa2,QAOA} are becoming
increasingly important for quantum computation. The main idea behind this
computational paradigm is to use hybrid classical-quantum algorithms that work
in tandem to solve computational problems.  The classical part of the algorithm
is performed by a classical processor and the quantum part of the algorithm is
executed on a quantum device. During the computation process, intermediary
results produced by the quantum device are passed onto the classical device
which performs further computation on them that is used to tune the parameters
of the quantum part of the algorithm, which therefore has an effect on the
quantum dynamics. The hybrid classical-quantum back and forth process repeats
until a desired termination condition is satisfied. 

This hybrid classical-quantum computational paradigm opens up interesting and
important challenges for the design of suitable programming languages.  It is
clear that if we wish to program within such computational scenarios, we need
to develop a language that correctly models the manipulation of \emph{quantum
resources}. In particular, quantum measurements give rise to
\emph{probabilistic computational effects} that are inherited by the classical
side of the language. Another issue is that quantum information behaves very differently compared to classical
information. As an example, quantum information cannot be copied in a uniform
way \cite{no-cloning}, unlike classical information, which may be freely copied
without restriction.  Therefore, if we wish to avoid runtime errors, the
quantum fragment of the language needs to be equipped with features for
fine-grained control, such as for example, having a \emph{substructural typing
discipline} \cite{girard,benton-small,benton-wadler,qtt1,qtt2} where
contraction (i.e.  copying) is restricted.  On the other hand, when doing
classical computation, such restrictions are unnecessary and often
inconvenient.  One solution to this problem is to design a language with a
classical (non-linear) fragment together with a quantum (linear) one, both of
which interact nicely with each other.  In fact, this can be achieved within an
existing language that has a sufficiently advanced type system, as we show in
this paper.
\subsubsection*{Our Contributions.}
In this paper, we describe \emph{\name{}} (named after the hybrid creature
Chimaera from Greek mythology), which is a set of libraries for the Idris 2
language \cite{idris2} that allow the programmer to implement
(variational) quantum algorithms in a \emph{type-safe} way. Idris 2
is an elegant functional programming language that is equipped with an advanced
type system based on Quantitative Type Theory \cite{qtt1,qtt2} that brings many
useful features to the programmer, most notably \emph{dependent types} and
\emph{linearity}. These two features of Idris are crucial for the development
of \name{} and, in fact, are the reason we chose Idris in the first place.
Dependent types are used throughout our entire development in order to
correctly represent and formalise the compositional nature of quantum
operations.  Linearity is used in order to enforce the proper consumption of
quantum resources (during execution) in a way that is admissible with respect
to the laws of quantum mechanics.  The combination of dependent types and
linearity allows us to \emph{statically} detect and reject erroneous quantum
programs and this ensures the type safety of our approach to variational
quantum programming.

In our intended computational scenario, we have access to both a
classical computer and a quantum computer. Since we cannot directly observe
quantum information, we directly interact with the classical computer which
sends instructions to, and receives data from, the quantum device via a
suitable interface that makes use of the IO monad. In our view, this is an
adequate representation of a realistic computational environment for
variational quantum programming.
We design a suitable (abstract) interface that allows us to model this
situation accurately and which makes use of the IO monad.
However, since the authors do not personally have any
quantum hardware, we provide only one concrete implementation of our interface that simulates the relevant quantum operations on our
classical computers by using the proper linear-algebraic formalism, but while
still using the IO monad as prescribed by the abstract interface. From a
high-level programming perspective, both the abstract interface and its
concrete implementation (via linear-algebraic simulation) address all of the
programming challenges induced by the classical-quantum device computational
scenario.

We emphasise that we can achieve type-safe (variational) quantum programming in
an \emph{existing} programming language by implementing suitable libraries.
This is important for \emph{variational} quantum programming, because in most
variational quantum algorithms, the classical part of the algorithm is
considerably larger, more complicated and more difficult to implement, compared
to the quantum part of the algorithm. Therefore, it is important for the
programming language to have first-class support for classical programming features.
Our chosen language, Idris, is definitely such a language. The advanced type
system of Idris allows us to elegantly mix quantum and classical programming
primitives and therefore allows us to get the best of both worlds.
We demonstrate that Qimaera
is suitable for variational quantum programming by providing implementations of
the two most prominent variational quantum algorithms -- QAOA and VQE. To the
best of our knowledge, this is the first implementation of these algorithms
that has been achieved in a \emph{type-safe} framework. In particular, this
means that common quantum programming errors (e.g. copying of qubits, applying a CNOT operation with the same source and target, etc.)
are \emph{statically} detected and rejected by the Idris type checker.

\subsubsection*{Overview and summary of contributions}
We begin by providing some background on quantum computation (\secref{sub:quantum-background}) and the Idris programming language (\secref{sub:idris-background}).
We then explain how we represent unitary (i.e. reversible and controllable) quantum operations in Idris and we provide some important and non-trivial examples (\secref{sec:quantum-unitaries}).
In \secref{sec:quantum-operations} we describe how we represent arbitrary (non-unitary, effectful) quantum operations and we present some examples of effectful quantum programs and algorithms. The linear features of Idris are crucial for achieving this.
We discuss why \name{} is suitable for variational quantum programming and we provide implementations of the VQE and QAOA algorithms in \secref{sec:variational}.
Finally, we discuss related work and future work in \secref{sec:related}.
The Idris source code for \name{} is available at \url{https://github.com/zamdzhiev/Qimaera} under the MIT license.
In this paper, we provide some excerpts of the source code, but we recommend consulting the code itself, because it contains many useful comments.

\section{Background}

In this section we introduce the relevant background so that readers may follow
our subsequent development.

\subsection{Quantum Computation}
\label{sub:quantum-background}

Readers interested in a detailed introduction to quantum computing may consult
\cite{nielsen-chuang}. In this section we summarise the basic notions that are
relevant for our development.

\subsubsection{Qubits}

The simplest non-trivial quantum system is the \emph{quantum bit}, often
abbreviated as \emph{qubit}. Qubits may be thought of as the quantum
counterparts of the bit from classical computation. A qubit $\ket \psi$ is represented as a normalised vector in $\mathbb C^2.$
The \emph{computational basis} is given by the pair of vectors 
$ \ket{0} \eqdef \begin{pmatrix} 1 \\ 0 \end{pmatrix}$  and  $\ket{1} \defeq \begin{pmatrix} 0 \\ 1 \end{pmatrix}  , $
which may be seen as representing the classical bits $0$ and $1$. 
An arbitrary qubit is described by $\ket \psi = a \ket 0 + b
\ket 1$ where $a,b \in \mathbb{C}$ and $|a|^2 + |b|^2 = 1$.

\subsubsection{Superposition}

A qubit may be in (uncountably) many different states, whereas a classical bit
is either $0$ or $1$. When the linear combination $\ket \psi = a \ket 0 + b
\ket 1$ is non-trivial, then we say that $\ket \psi$ is in \emph{superposition}
of $\ket 0$ and $\ket 1$.
Superposition is a very important quantum resource
which is used by many quantum algorithms.

\subsubsection{Composite Systems}

The state space that describes a system of $n$ qubits is the Hilbert space
$\mathbb{C}^{2^n}$. 
If $\ket{\psi}$ and $\ket{\phi}$ are
two states of $n$ and $m$ qubits respectively, then the composite $n+m$ qubit
state $\ket{\psi\phi} \eqdef \ket{\psi} \otimes \ket{\phi}$ is described by the
Kronecker product $\otimes$ of the original states.

\subsubsection{Unitary Quantum Operations}
\label{sub:unitary-ops}
A quantum state $\ket \psi \in \mathbb C^{2^n}$ may undergo a \emph{unitary
evolution} described by a unitary matrix $U \in \mathbb C^{2^n \times 2^n}$ in
which case the new state of the system is described by the vector $U \ket
\psi.$ Unitary operations (and matrices) are closed under sequential composition
(described by matrix multiplication $\circ$) and under parallel composition (described
by Kronecker product $\otimes$ ). Sequential composition of unitary operations is used to
describe the temporal evolution of quantum systems, whereas the parallel
composition is used to describe their spatial structure.

The unitary quantum operations are also often called \emph{unitary gates}.  One
typically chooses a \emph{universal gate set} which is a small set of unitary
operations that suffices to express all other unitary operations via (parallel
and sequential) composition. The universal gate set that we choose for our
development is standard and we specify these unitary operations next by
giving their action on the computational basis (which uniquely determines
the operations).

The \emph{Hadamard Gate}, denoted $H$, is the 1-qubit unitary map whose action on the computational basis is given by
\[ H \ket{0} = \frac{1}{\sqrt{2}} (\ket{0} + \ket{1}) \hspace{0.5cm} H \ket{1} = \frac{1}{\sqrt{2}} (\ket{0} - \ket{1})  \]
and its primary purpose is to generate superposition.
The \emph{Phase Shift Gate}, denoted $P(\alpha)$, for $\alpha \in \mathbb R$, is a 1-qubit unitary map whose action on the computational basis is given by:
\[ P(\alpha) \ket{0} = \ket{0} \hspace{0.5cm} P(\alpha) \ket{1} = e^{i \alpha} \ket{1} \]
and its primary purpose is to modify the phase of a quantum state. 
The family of Phase Shift Gates is parameterised by the choice of $\alpha \in \mathbb R$ and important special cases include the unitary gates $T \eqdef P(\pi /4)$ and $Z \eqdef P(\pi).$
The \emph{Controlled-Not Gate} (CNOT), is a 2-qubit unitary map whose action on the computational basis is given by
\begin{align*}
 \mathrm{CNOT} \ket{00} = \ket{00} \quad \mathrm{CNOT} \ket{01} = \ket{01} \\
 \mathrm{CNOT} \ket{10} = \ket{11} \quad \mathrm{CNOT} \ket{11} = \ket{10}
\end{align*}
and this unitary map may be used to generate quantum entanglement (see \secref{subsub:entanglement}).

\begin{figure}
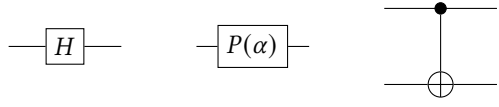

  \centering
  \stikz{atomic-gates.tikz}
  \caption{The Hadamard, Phase Shift and CNOT gates.}
  \label{fig:gates}
\end{figure}


Unitary gates admit a diagrammatic representation as \emph{quantum circuits}. The atomic unitary gates we described above are shown in Figure \ref{fig:gates}.
Composite unitary gates may also be described as circuits (see Figure \ref{fig:bell-state}): sequential composition amounts to plugging wires of subdiagrams and parallel composition amounts to juxtaposition.

\subsubsection{Controlled Unitary Operations}

The $\mathrm{CNOT}$ gate is the simplest example of a \emph{controlled unitary gate}. Given a unitary gate $U \colon \mathbb C^{2^n} \to \mathbb C^{2^n}$,
the controlled-$U$ unitary gate is the unitary gate $CU \colon \mathbb C^{2^{n+1}} \to \mathbb C^{2^{n+1}} $ whose action is determined by the assignments
\[ CU(\ket 0 \otimes \ket \psi) = \ket{0} \otimes \ket \psi \qquad CU(\ket 1 \otimes \ket \psi) = \ket 1 \otimes (U \ket \psi) . \]
Controlled unitary operations are ubiquitous in quantum computing and they are  graphically depicted as
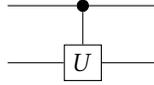
\cstikz{controlled-unitary.tikz}
using a similar notation to that of the $\mathrm{CNOT}$ gate.

\subsubsection{Inverse Unitary Operations}

Every unitary operation $U$ is \emph{reversible} with the inverse operation
given by the conjugate transpose, denoted $U^\dagger$, which is again a unitary
matrix. Applying the inverse operation (also known as the \emph{adjoint}) of a
given unitary is also ubiquitous.

\subsubsection{Quantum Entanglement}
\label{subsub:entanglement}

A quantum state $\ket \psi \in \mathbb C^{2^n}$, with $n > 1,$ is said to be
\emph{entangled} when there exists no non-trivial decomposition $\ket \psi =
\ket \phi \otimes \ket \tau$. Quantum entanglement is a very important resource in
quantum computation which is exhibited by many quantum algorithms. Because of
the possibility of entanglement, we cannot, in general, break down quantum
systems into smaller components and we are often forced to reason about such
systems in their entirety.
The most important example of an entangled state is the \emph{Bell state} given by $\ket{\mathrm{Bell}} \eqdef \frac{\ket{00} + \ket{11}}{\sqrt{2}}$.

\subsubsection{Preparation of Quantum States}

\begin{figure}
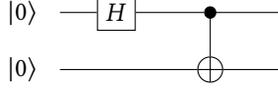

  \centering
  \stikz{bell.tikz}
  \caption{Preparation of the Bell state using atomic gates.}
  \label{fig:bell-state}
\end{figure}

Preparing a new qubit in state $\ket 0$ is an admissible physical operation.
This, together with application of unitary gates as part of the computation,
allows us to prepare arbitrary quantum states, e.g., the Bell state can
be prepared by taking $\ket{\mathrm{Bell}} = (\mathrm{CNOT} \circ (H \otimes
I)) \ket{00}$. See Figure~\ref{fig:bell-state} for the corresponding circuit.

\subsubsection{Measurements}
\label{subsub:measurements}
Quantum information cannot be directly observed without affecting the state of
the underlying system. In order to extract information from quantum systems, we
need to perform a \emph{quantum measurement} on (parts of) our systems.  For
example, when performing a quantum measurement on a qubit in the state
$\ket{\psi} = a \ket{0} + b \ket{1}$, there are two possible outcomes: either
the quantum system will collapse to state $\ket 0$ and we obtain the classical
bit $0$ as evidence of this event, or, the quantum system will collapse to
state $\ket{1}$ and we obtain the classical bit $1$ as evidence of this event.
The first outcome (corresponding to bit $0$) occurs with probability $|a|^2$
and the second outcome (corresponding to bit $1$) occurs with probability $1 -
|a|^2= |b|^2.$ In general, when we measure $n$ qubits simultaneously, we obtain
a bit string of length $n$ which determines the event that occurred and the
quantum system collapses to a corresponding state with some probability, both
of which are determined via the Born rule of quantum mechanics.
Therefore, quantum measurements induce evolutions which are \emph{probabilistic} and \emph{irreversible} (or \emph{destructive}),
which distinguishes them from unitary evolutions, which are \emph{deterministic} and \emph{reversible}.

\subsubsection{No-Cloning Theorem}
Unlike classical information, quantum information cannot be uniformly copied.
This is made precise by the \emph{no-cloning} theorem of quantum mechanics
\cite{no-cloning}: there exists no unitary operation $U : \mathbb{C}^4
\rightarrow \mathbb{C}^4$, such that for every qubit $\ket{\psi} $ :
$ U (\ket{\psi} \otimes \ket{0}) = \ket{\psi} \otimes \ket{\psi} . $
This means that copying of quantum information is a \emph{physically inadmissible}
operation. Ideally, quantum programming languages should be designed so that
these kinds of errors are detected during type checking
and not at runtime.

\subsection{The Idris 2 Language}
\label{sub:idris-background}
In this section, we give a short overview of the Idris 2 language
and its main features that are relevant for the development of Qimaera.  Idris
2 is a pure functional language with a syntax influenced by that of Haskell.
The main additional features, compared to Haskell, are dependent types and
linearity, both of which are crucial for \name{}. Its type system is based on
Quantitative Type Theory \cite{qtt1,qtt2}, which specifies how dependent types
and linearity are combined.

\subsubsection{Dependent Types}
In Idris, types are first-class primitives and they may be manipulated like any
other construct of the language. This allows us to formulate more expressive
types that can depend on values, and hence it enables us to make some
properties and program invariants explicit.

\begin{example}
\label{ex:vector}
  The type of vectors is a simple and useful example of a dependent type. A
  vector is a list with a fixed length that is part of the type. It can be
  defined as follows, where \texttt{S} is the successor function for
  natural numbers, and \texttt{a} is a polymorphic type: 
\begin{lstlisting}
data Vect : Nat -> Type -> Type where
  Nil  : Vect 0 a 
  (::) : a -> Vect k a -> Vect (S k) a
\end{lstlisting}
  The type \texttt{Vect} has two constructors (i.e., introduction rules). The
  first one constructs the empty vector, of length zero. The second one is used
  to introduce non-empty vectors: a vector with \texttt{k+1} elements of type
  \texttt{a} is constructed by combining an element of type \texttt{a} and a
  vector of size \texttt{k}.
\end{example}

Type dependency allows us to specify useful program properties and type
checking ensures that they hold. For instance, we can define an
\texttt{append} function that concatenates two vectors. Then, the
size of the output vector is the sum of the sizes of the input vectors and this
is specified by its type.
\begin{lstlisting}[basicstyle=\fontsize{8.5}{9}\ttfamily]
append : Vect n a -> Vect m a -> Vect (n + m) a
\end{lstlisting}
This information allows the language to detect a larger class of programming errors.
%
Note that type dependency information is not available for the analogous
function on lists.
Type dependency may also be used to express constraints on the inputs of a
function, e.g. we can define a \emph{total} function, called \texttt{pop},
that cannot be applied to an empty vector. 
\begin{lstlisting}
pop : Vect (S k) a -> Vect k a
pop (x :: xs) = xs
\end{lstlisting}
Writing "\texttt{pop []}" is now an error which is detected statically, rather
than dynamically, and we note that the same cannot be achieved if we were to
replace vectors with lists.

With the addition of dependent types, we sometimes have to prove the
equivalence between types. For example, the types \texttt{Vect (n + k) Nat} and
\texttt{Vect (k + n) Nat} are equivalent, but Idris cannot (at present) infer
this, so the programmer has to sometimes provide a proof that addition is
commutative (which is easy to do). All proof obligations needed for
our development are simple and concise.

\subsubsection{Linearity}

The type system of Idris 2 is based on Quantitative Type Theory, where every
function argument is associated with a multiplicity that states the number of
times the variable is used at runtime. This multiplicity can be 0, 1 or
$\omega$. An argument with multiplicity $0$ is only used at compile time (to
determine type dependency information) and is erased at runtime.  A
\emph{linear} argument has multiplicity 1 and it is used exactly once at
runtime. Finally, $\omega$ represents the unrestricted multiplicity, which is
also default, where the function argument may be used any number of times.

\begin{example}
  Consider the \texttt{pop} function which we just discussed. The (implicitly
  bound) variables \texttt{k} and \texttt{a} have multiplicity 0, because they
  are not explicitly specified as separate arguments, and they are \emph{not}
  accessible at runtime in the function. The variables \texttt{x} and
  \texttt{xs}, which are explicitly bound, have the default (unrestricted) multiplicity.
\end{example}

\begin{example}
  An important type which we define in \name{} is the type of linear vectors,
  which we write as \texttt{LVect}. The only difference, compared to the standard vectors in Idris, is that the
  \texttt{(::)} constructor for \texttt{LVect} is a linear function in all of its arguments.
  Linearity in Idris 2 is specified by writing the multiplicity $1$ in front of each argument. 
\begin{lstlisting}
data LVect : Nat -> Type -> Type where
  Nil : LVect 0 a
  (::) : (1 _ : a) -> (1 _ : LVect k a) -> 
         LVect (S k) a
\end{lstlisting}
  We also use linear pairs that are already defined in Idris 2. 
\begin{lstlisting}
data LPair : Type -> Type -> Type
  (#) : (1 _ : a) -> (1 _ : b) -> LPair a b
\end{lstlisting}
\end{example}

Linearity allows us to specify and enforce constraints on function
arguments, e.g. it prevents us from duplicating data, so the function
definition below leads to an error:
\begin{lstlisting}
copy : (1 _ : a) -> LPair a a
copy x = x # x

Error: While processing right hand side of 
copy. There are 2 uses of linear name x.
\end{lstlisting}

Linearity is prominently used in Qimaera. In particular, when manipulating
quantum information, linearity is enforced in order to properly consume quantum
resources and comply with the laws of quantum mechanics.

\section{Unitary Operations in \name{}}
\label{sec:quantum-unitaries}
As we saw in \secref{sub:quantum-background}, unitary transformations have a
special role in quantum computation. In most non-variational quantum
algorithms, the vast majority of the programming effort consists in
implementing the required unitary operations. In this section, we describe our
representation of unitary transformations in \name{} as an algebraic data type
called \texttt{Unitary}. Every value of this type is, by design, an \emph{algebraic decomposition}
of a unitary operation in terms of the atomic unitary gates that we identified in \secref{sub:unitary-ops}.

The \texttt{Unitary} data type is very useful, because it allows us to adopt a
high-level \emph{algebraic} and \emph{scalable} approach towards the reversible fragment of quantum
computation.  This provides the programmer with many benefits as we show in
this section.  However, using the \texttt{Unitary} data type is actually
entirely optional.  Users who are interested in effectful quantum programming
do \emph{not} have to use it (see \secref{sec:quantum-operations}) and they may
still recover the full power of (variational) quantum programming, but at the
cost of losing the algebraic decomposition of unitary operations.  To utilise
the full potential of \name{}, programmers should make use of all the
constructions described in this section and the next one.

\subsection{The Unitary Data Type}

Quantum unitary operations admit a compositional and algebraic representation
based on the atomic gates from the universal gate set in \secref{sub:unitary-ops}.
Our idea for the representation of unitary operations is based on this, or
equivalently, on how unitary operations may be expressed in terms of unitary
quantum circuit diagrams.  Because of these reasons, linearity is not required
for our formalisation of unitary operations.  The code for the \texttt{Unitary}
data type is listed in Figure~\ref{fig:unitary} and we now describe our
representation in greater detail. 

\begin{figure}
\centering
\begin{lstlisting}
data Unitary : Nat -> Type where
  IdGate : Unitary n
  H      : (j : Nat) -> 
           {auto prf : (j < n) = True} -> 
           Unitary n -> Unitary n
  P      : (p : Double) -> (j : Nat) -> 
           {auto prf : (j < n) = True} -> 
           Unitary n -> Unitary n
  CNOT   : (c : Nat) -> (t : Nat) -> 
           {auto prf1 : (c < n) = True} -> 
           {auto prf2 : (t < n) = True} -> 
           {auto prf3 : (c /= t) = True} -> 
           Unitary n -> Unitary n
\end{lstlisting}
\caption{The Unitary data type (file: \texttt{Unitary.idr}).}
\label{fig:unitary}
\end{figure}

Given a natural number \texttt{n : Nat}, the type of unitary operations
on \texttt{n} qubits is given by \texttt{Unitary n}. Note that \texttt{Unitary}
is an algebraic data type with a simple type dependency on the arity of the
desired operation.  The \texttt{Unitary} type has four different introduction
rules which we describe next. 

The first constructor,\ \texttt{IdGate}, represents the identity unitary
operation on \texttt{n} qubits. Diagramatically, we can see this as
constructing a circuit of \texttt{n} wires, without applying any other gates on
any of the wires. It has a unique argument, \texttt{n}, which is implicit -- it
can be omitted when calling the \texttt{IdGate} constructor and it will often
be inferred by Idris.

The second constructor, \texttt{H}, should be understood as applying the Hadamard
gate $H$ to the \texttt{j}-th qubit of some previously constructed unitary circuit
which is specified as the last argument. The first implicit argument,
\texttt{n}, is simply the arity of the resulting unitary operation. The second
implicit argument, \texttt{prf}, is a proof obligation that \texttt{j} is
smaller than \texttt{n}. This ensures that the argument \texttt{j} identifies
an existing wire of the previously constructed unitary circuit (last argument) and
therefore the overall definition is algebraically and physically sound. We
note that the implicit argument \texttt{prf} may be removed from our
implementation if we change the type of \texttt{j} to \texttt{Fin n},
the type of integers less than \texttt{n}. However,
in our experience, Idris has better support for \texttt{Nat} than for
\texttt{Fin} and for this reason we chose to keep the \texttt{prf} argument.

The third constructor, \texttt{P}, should be viewed as applying the
$P(p)$ gate, where the real number $p \in \mathbb R$ is approximated by the
term \texttt{p : Double}.\footnote{This approximation is not a big
limitation -- in fault-tolerant quantum computing one usually replaces the
$P(p)$ gate family with a single $T = P(\pi/4)$ gate and the resulting gate set
suffices to approximate any unitary with \emph{arbitrary} precision. So we can easily replace \texttt{P} with a \texttt{T} constructor.} The remaining
arguments serve the same purpose as those for \texttt{H}.

The final constructor, \texttt{CNOT}, should be understood as applying the
$\mathrm{CNOT}$ gate, where \texttt{c} identifies the wire used for the control
(the small black dot in Figure \ref{fig:gates}), \texttt{t} identifies the wire
of the target (the crossed circle in Figure \ref{fig:gates}) and the last
(unnamed) argument is the previously constructed unitary circuit on which we
are applying CNOT. The remaining arguments are implicit and often do not have
to be provided by the users: the argument \texttt{n} is the arity of the
unitary; \texttt{prf1} and \texttt{prf2} ensure that \texttt{c} and \texttt{t}
identify valid wires of the unitary circuit; \texttt{prf3} ensures that the
control and target wires are \emph{distinct} and therefore the overall
application of CNOT is physically and algebraically admissible.

In our representation of quantum unitary operations, we make use of
type dependency to impose proof obligations on some of our
constructors in order to guarantee that the representation makes sense in
physical and algebraic terms. On first glance, this might seem like a burden
for the users of the library. However, in our experience, Idris can often
automatically infer the required proofs (without any assistance from the user)
and we had to do very little manual theorem proving. This is discussed in
detail in the next subsection.

\subsection{Constructing Unitary Transformations}

The four basic introduction rules of the \texttt{Unitary} type allow us to define \emph{high-level functions} in
Idris that can be used to construct complex unitary circuits out of simpler ones.
We discuss this here and we show that the proof obligations from Figure
\ref{fig:unitary} are not severe and can be easily ameliorated and often
completely sidestepped using our high-level functions.

First, we point out that auto-implicit arguments may often be inferred by Idris
via suitable search. For example, if all the arguments are known
statically, the required proofs will be discovered by Idris and the users do
not have to manually provide them.

\begin{example}
\label{ex:bell-unitary}
The unitary gate depicted in the circuit from Figure \ref{fig:bell-state} may
be constructed in the following way:
\begin{lstlisting}
toBellBasis : Unitary 2
toBellBasis = CNOT 0 1 (H 0 IdGate)
\end{lstlisting}
In this example, Idris is able to infer all the implicit arguments and there is no need to provide any proofs.
If we do not satisfy one of the constraints, e.g. if we write \texttt{CNOT 1 1} above (which does not make physical sense),
then we get the following error during type checking:
\begin{lstlisting}
Error : While processing right hand side of
toBellBasis. Can't find an implementation for
not (== 1 1) = True.
\end{lstlisting}
An error is also reported if we provide a wire number larger than 1.
It is also useful to define \emph{standalone} unitary gates for the $H, P(r)$ and $\mathrm{CNOT}$ gates as follows:
\begin{lstlisting}[basicstyle=\fontsize{8.5}{9}\ttfamily]
HGate : Unitary 1   
HGate = H 0 IdGate  

PGate : Double -> Unitary 1
PGate r = P r 0 IdGate

CNOTGate : Unitary 2
CNOTGate = CNOT 0 1 IdGate
\end{lstlisting}
\end{example}

\subsubsection{Composing Unitary Circuits}

Our libraries provide functions for sequential composition (\texttt{compose}) and parallel composition (\texttt{tensor}) of unitary operations:
\begin{lstlisting}[basicstyle=\fontsize{8.5}{9}\ttfamily]
compose : Unitary n -> Unitary n -> Unitary n
tensor : {n : Nat} -> {p : Nat} -> Unitary n -> Unitary p -> Unitary (n + p)
\end{lstlisting}
Notice that both functions do not impose any proof obligations on the user.
This means that the \emph{primary} algebraic way for composing unitary
operations may be done without \emph{any} need for theorem proving.
The use of these functions is ubiquitous in practice and we introduce the infix synonyms
\texttt{(.)} and \texttt{(\#)} for \texttt{compose} and \texttt{tensor}, respectively.

\begin{example}
The \texttt{toBellBasis} gate from Example \ref{ex:bell-unitary} may be equivalently expressed in the following way:
\begin{lstlisting}
toBellBasis : Unitary 2
toBellBasis = CNOTGate . (HGate # IdGate)
\end{lstlisting}
\end{example}

\name{} provides another, more general, form of composition via the function \texttt{apply} whose type is
as follows:
\begin{lstlisting}
apply : {i : Nat} -> {n : Nat} -> 
        Unitary i -> Unitary n -> 
        (v : Vect i Nat) -> 
        {auto _ : isInjective n v = True} ->
        Unitary n
\end{lstlisting}
The \texttt{apply} function is used to apply a smaller unitary circuit of size
\texttt{i} to a bigger one of size \texttt{n}, giving the vector \texttt{v} of
wire indices on which we wish to apply the smaller circuit. It needs one
auto-implicit proof which enforces the consistency requirement that all indices
of the wires specified by \texttt{v} are pairwise distinct and smaller than
\texttt{n}.
In fact, the \texttt{apply} function implements the most general notion of
composition possible and both sequential and parallel composition can be realised as
special cases using it. The importance of the vector \texttt{v} is that it
determines how to apply the smaller unitary circuit of arity \texttt{i} to
\emph{any selection} of \texttt{i} wires of the larger unitary circuit, and
moreover, it also allows us to \emph{permute} the inputs/outputs of the smaller
unitary circuit while doing so. More specifically, if the $k$-th entry of the
vector \texttt{v} is the natural number $p$, then the $k$-th input/output of
the smaller unitary circuit will be applied to the $p$-th wire of the larger
unitary circuit. This is best understood by example.

\begin{example}
Consider the following code sample:
\begin{lstlisting}
U : Unitary 3
U = HGate # IdGate {n = 1} # (PGate pi)

apply_example : Unitary 3
apply_example = apply toBellBasis U v
\end{lstlisting}
  where \texttt{v} is a vector of length two.
  Here \texttt{toBellBasis} is given in Example \ref{ex:bell-unitary} and represents the circuit given below left;
  \texttt{U} represents the circuit given below right:
  \[ \stikz{simple-examples.tikz} \]
  Table \ref{tb:apply} shows what
  unitary circuit is specified under different values of \texttt{v}.  
  Here, Idris can automatically infer the required proofs and the user does
  not have to provide them.
\begin{table}
\centering
  \begin{tabular}{|c|c|}
  \hline 
    \texttt{apply toBellBasis U [0,1]} & \stikz{apply-example1.tikz} \\
  \hline
    \texttt{apply toBellBasis U [0,2]} & \stikz{apply-example2.tikz} \\
  \hline
    \texttt{apply toBellBasis U [2,0]} & \stikz{apply-example3.tikz} \\
  \hline
    \texttt{apply toBellBasis U [2,1]} & \stikz{apply-example4.tikz} \\
  \hline
  \end{tabular}
  \caption{Examples illustrating the \texttt{apply} function.}
  \label{tb:apply}
\end{table}
\end{example}

\begin{remark}
  Instead of using \texttt{apply}, there is another possible approach, in the
  spirit of \emph{symmetric monoidal categories} \cite[\S XI]{maclane}, where we could add one extra
  introduction rule to the \texttt{Unitary} type for representing
  \emph{permutations} of wires. However, in our view, this approach is somewhat
  awkward, because one does not usually think of permutations (induced by the
  symmetric monoidal structure) as physical gates.
\end{remark}

\subsubsection{Adjoints of Unitary Circuits}

\name{} also provides a function
\begin{lstlisting}
adjoint : Unitary n -> Unitary n
\end{lstlisting}
which computes the adjoint (i.e., inverse) of a given unitary circuit. As
explained previously, one often has to apply the inverse of a given unitary
circuit, so having a high-level method such as this is useful. Our
implementation uses the obvious algorithm for synthesising the adjoint. This
may be used, for example, to automatically uncompute operations that we perform
on ancilla qubits, which is often required.

\subsubsection{Controlled Unitary Circuits}
\label{sub:controlled}
We also implement a function
\begin{lstlisting}
controlled : {n : Nat} -> Unitary n -> Unitary (S n)
\end{lstlisting}
which given a unitary circuit $U$ constructs the corresponding controlled
unitary circuit $CU$. Our implementation uses the obvious algorithm for doing this,
but more efficient algorithms may also be implemented in the future.

\subsubsection{Analysis of Unitary Circuits}

Unitary circuits are represented in a scalable and compositional way in
\name{}, thus we can use Idris to optimise them. The function:
\begin{lstlisting}
optimise : Unitary n -> Unitary n
\end{lstlisting}
may be used to optimise a given (very large) unitary circuit. So far, this
function provides some basic optimisations, but more sophisticated ones may be
added in the future. The point we wish to make is that unitary circuits in
\name{} may be analysed and manipulated like any other algebraic data type
using the full capabilities of Idris. In fact, the file \texttt{Unitary.idr}
provides many functions that do this, e.g. we provide functions for calculating
the circuit depth, calculating the number of specific atomic gates used by a
circuit, drawing circuits in the terminal, exporting circuits to Qiskit
(so that users may then use external analysis tools), etc.

\subsection{Example: The Quantum Fourier Transform}

\begin{figure}
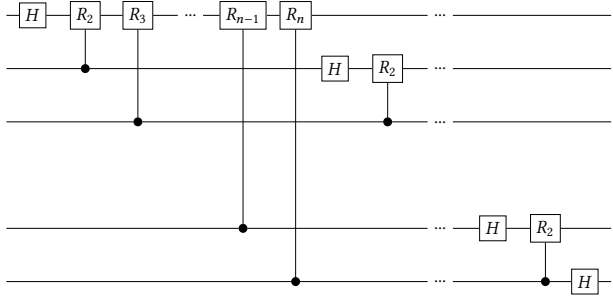

  \centering
  \stikz[0.7]{qft.tikz}
  \caption{The QFT unitary circuit on $n$ qubits.}
  \label{fig:qft}
\end{figure}

The Quantum Fourier Transform (QFT) is a very important unitary operator that
is used in Shor's polynomial-time algorithm for integer factorisation
\cite{shor}.  The unitary circuit which realises QFT on $n$ qubits is shown in
Figure ~\ref{fig:qft}, where $R_n \eqdef P \left( \frac{2\pi}{2^n} \right).$
The \name{} code which implements this unitary circuit is shown in
Figure~\ref{fig:qftCode}. Notice that we make use of the \texttt{controlled}
function from \secref{sub:controlled} in the function \texttt{cRm}, so that we
can automatically implement the many controlled $R_n$ gates that are required.
In this example, all the parameters are universally quantified, so we need a
few very short and simple proofs in the code: one for using the \texttt{apply}
function and one for correctly unifying the size of the circuit.  Currently
Idris cannot automatically discover these proofs, but we hope in the future its
capabilities for proof search would improve to the point where it could. If
this were to happen, then we would not have to manually provide these proofs.
\begin{figure}
\begin{lstlisting}
Rm : Nat -> Unitary 1
Rm m = PGate (2 * pi / (pow 2 (cast m)))

cRm : Nat -> Unitary 2
cRm m = controlled (Rm m)

qftRec : (n : Nat) -> Unitary n
qftRec 0 = IdGate
qftRec 1 = HGate
qftRec (S (S k)) =
  let t =  (qftRec (S k)) # IdGate
  in rewrite sym $ lemmaplusOneRight k 
  in apply (cRm (S k)) t [S k,0] 
           {prf = lemmaInj1 k}
           
qft : (n : Nat) -> Unitary n
qft 0 = IdGate
qft (S k) =
  let g = qftRec (S k)
      h = (IdGate {n = 1}) # (qft k)
  in h . g
\end{lstlisting}
  \caption{\name{} code for QFT (file: \texttt{QFT.idr}).}
  \label{fig:qftCode}
\end{figure}

\section{Effectful Quantum Computation}
\label{sec:quantum-operations}
In the previous section we showed how unitary circuits can be represented in
\name{} in a compositional, algebraic and scalable way.  This suffices to
capture the pure, deterministic and reversible fragment of quantum computation.
However, as we explained in \secref{sub:quantum-background}, we need to also
consider effectful and probabilistic quantum processes (e.g.  measurements) in
order to recover the full power of quantum computation. In this section we show
how this can be done in a type-safe way by using monads, linearity and
dependent types.


\subsection{Representation of Quantum Effects}

We now explain how the quantum program dynamics are represented in \name{} in a type-safe way.
We are (roughly) inspired by representing the notion of a \emph{quantum
configuration} as it appears in \cite{quantum-lambda,qpl-fossacs,vqpl}, which is in turn used
to formally describe the operational semantics of quantum type systems.

\subsubsection{Qubits in \name{}}

Because of the possibility of quantum entanglement (see
\secref{subsub:entanglement}), we cannot describe the state of an individual
qubit which is part of a larger composite system.  On the other hand, we wish
to be able to refer to \emph{parts} of the whole system by identifying specific
qubit positions.  In \name{}, we introduce the following type declaration:
\begin{lstlisting}
data Qubit : Type where
  MkQubit : (n : Nat) -> Qubit
\end{lstlisting}
The argument of type \texttt{Nat} is used as a \emph{unique identifier} for the
constructed qubit.  The constructor \texttt{MkQubit} is \emph{private} and
users of our libraries cannot access it. Instead, our libraries provide
functions (discussed later) that ensure that terms of type \texttt{Qubit} are
always created with a fresh (i.e. unique) natural number that serves as its
identifier. In fact, these functions are the only way users can access or
manipulate qubits and, moreover, our users cannot access these unique
identifiers.  This allows us to formulate a representation where terms of type
\texttt{Qubit} unambiguously refer to the relevant parts of larger composite
systems. Therefore, a term of type \texttt{Qubit} should be understood as a
pointer, or as a unique identifier, of a 1-qubit subsystem of some larger
quantum state. Terms of type \texttt{Qubit} should \emph{not} be understood as
representing any sort of state, because they do not carry such information.

\subsubsection{Probabilistic Effects}

As we previously discussed in \secref{subsub:measurements}, quantum measurements induce
probabilistic computational effects which are inherited by the classical side
of the computation in variational quantum algorithms. Furthermore, in our
intended computational scenario, the classical computer (on which Idris is
running) sends instructions to, and receives data from, the quantum device.
In order to correctly model all of this, it is clear that we have to use the IO monad in order to encapsulate these effects.

However, when representing quantum program dynamics, we also need to
\emph{enforce linearity}, but all the functions provided by the IO monad (e.g.
\texttt{pure} which introduces pure values to monadic types) are \emph{not} linear in any of their arguments. This creates a
problem which may be solved by using the LIO library, which extends the IO
monad with linearity. For brevity, we define \texttt{R} to be our
linear IO monad:
\begin{lstlisting}
R : Type -> Type
R = L IO {use = Linear}
\end{lstlisting}
Then, by using \texttt{R} we can combine IO effects (and thus also
probabilistic effects) and linearity in a suitable way.

\subsubsection{Quantum State Transformer}

Quantum computation is \emph{effectful}, and moreover, quantum information \emph{cannot} be observed by the classical
computer (on which Idris is running): it only receives classical information through communication with
the quantum device.  Because of this, we have to adopt a more abstract view on
the hybrid classical-quantum computational process.  In order to do this, we
define an (abstract) \emph{quantum state transformer} by combining several different
concepts: \emph{indexed state monads} \cite{indexed-monads}\footnote{See
\cite{indexed-monads-haskell} for a Haskell implementation of this idea.},
linearity and IO (and thus also probabilistic) effects. Our representation of
these ideas in \name{} is shown in Figure~\ref{fig:quantumOp}, where we omit
the function definitions for brevity.

\begin{figure}
\begin{lstlisting}[basicstyle=\fontsize{8.5}{9}\ttfamily]
data QStateT :
 Type -> Type -> Type -> Type where
  MkQST : 
    (1 _ : (1 _ : initialType) -> 
            R (LPair finalType returnType)) -> 
    QStateT initialType finalType returnType

runQStateT : (1 _ : initialType) ->
             (1 _ : QStateT initialType 
                    finalType returnType) ->
             R (LPair finalType returnType)

pure : (1 _ : a) -> QStateT t t a

(>>=) : (1 _ : QStateT i m a) ->
        (1 _ : ((1 _ : a) -> QStateT m o b)) ->
        QStateT i o b
\end{lstlisting}
  \caption{Quantum state transformer (file: \texttt{QStateT.idr}).}
\label{fig:quantumOp}
\end{figure}

The type \texttt{QStateT} is parameterised by a choice of three (arbitrary)
types, so it is fairly abstract. Soon, we will see that it is very useful for
our purposes. The intended interpretation of this type is the following: any
value of type
\[ \texttt{QStateT initialType finalType returnType} \]
represents a
stateful (quantum) computation starting from a (quantum) state of type
\texttt{initialType} and ending in a (quantum) state of type \texttt{finalType}
which produces a user-accessible result of type \texttt{returnType} during
the computation. For example, a value of type
\[ \texttt{QStateT (LPair Qubit Qubit) Qubit Bool} \]
should be understood as a quantum process that transforms a two-qubit state
into a single-qubit state and returns a single (classical) value of type
\texttt{Bool} to the user. The functions presented in Figure \ref{fig:quantumOp}
allow us to adopt a \emph{monadic programming discipline} when working with
\texttt{QStateT} and we do so henceforth. We remark \texttt{QStateT}
makes use of the monad \texttt{R} which encapsulates the IO (and
probabilistic) effects. Note also that linearity is enforced when working with \texttt{QStateT}.

\subsubsection{Effectful Quantum Programming}

The \texttt{QStateT} monad can be used to define a suitable \emph{abstract
interface} for quantum programming.  In Figure \ref{fig:quantum-operations}, we
present an excerpt of the \texttt{QuantumOp} interface which allows us to
easily write quantum programs and execute them in a type-safe way. All of the
(variational) quantum algorithms we present are implemented using this
interface that we describe next.

\begin{figure}
\begin{lstlisting}[basicstyle=\fontsize{8.5}{9}\ttfamily]
interface QuantumOp (0 t : Nat -> Type) where
    newQubits : (p : Nat) -> 
      QStateT (t n) (t (n+p)) (LVect p Qubit)
             
    applyUnitary : {n : Nat} -> {i : Nat} ->
      (1 _ : LVect i Qubit) -> Unitary i -> 
      QStateT (t n) (t n) (LVect i Qubit)
               
    measure : {n : Nat} -> {i : Nat} ->
      (1 _ : LVect i Qubit) ->
      QStateT (t (i + n)) (t n) (Vect i Bool)

    run : QStateT (t 0) (t 0) (Vect n Bool) ->
      IO (Vect n Bool)
\end{lstlisting}
  \caption{The \texttt{QuantumOp} interface (file: \texttt{QuantumOp.idr}).}
\label{fig:quantum-operations}
\end{figure}

The function \texttt{newQubits} is used to prepare \texttt{p} new qubits in
state $\ket 0$ and the function returns a linear vector of length \texttt{p}
with the qubit identifiers of the newly created qubits.

The function \texttt{applyUnitary} is used to apply a unitary operation of
arity \texttt{i} to the qubits specified by the argument \texttt{LVect} (which
also determines the order of application) and the operation returns an
\texttt{LVect} which serves the same purpose -- it identifies the qubits which
were just modified by the unitary operator.
The file \texttt{QuantumOp.idr} also provides functions \texttt{applyH},
\texttt{applyP} and \texttt{applyCNOT} which can be seen as special cases of
\texttt{applyUnitary}. However, these three functions do not depend on the
\texttt{Unitary} type.

The \texttt{measure} function is used to measure \texttt{i} qubits identified
by the \texttt{LVect} argument and it returns a value of type \texttt{Vect i
Bool} that represents the result of the measurement. During this process, the
\texttt{i} qubits are destroyed, as one can see from the provided type information.

Finally, the function \texttt{run} is used to \emph{execute} quantum algorithms
on the quantum device and obtain the classical information returned from it.
Notice that \texttt{run} can be used to execute effectful quantum processes
which start from the trivial quantum state (on zero qubits) and which terminate
in the same trivial quantum state, but which also produce some number of
classical bits as a user-accessible return result. This may be used to run
quantum algorithms: in a typical situation, we start with the trivial quantum
state (on zero qubits), we prepare $n$ qubits in state $\ket 0$, we apply some
unitary operations on them, and we finally measure all the qubits, thereby
destroying all the qubits and producing $n$ bits of classical information. This
quantum algorithm may then be represented as a value of type \texttt{QStateT (t
0) (t 0) (Vect n Bool)}.  Running it, however, produces a classical value of
type \texttt{IO (Vect n Bool)}, because the execution is probabilistic and
because our classical computer (on which we are running Idris) has to perform
IO actions to communicate with the quantum device.

In fact, \emph{all} of the above operations modify the quantum state
on the quantum device and may cause IO effects, because of the need to
communicate with the quantum device. This is indeed reflected by our
implementation. Observe, that our interface is defined using the
\texttt{QStateT} monad transformer which does incorporate IO effects (via the
\texttt{R} monad we discussed previously).

The best way to understand how this interface may be used for quantum
programming is to look at examples.

\begin{example}
\label{ex:coin}
  A fair coin toss may be implemented using quantum resources. The process is
  simple: (1) prepare the state $\ket 0$; (2) apply the $H$ gate to it; (3)
  measure the qubit and return this as output. We implement this as follows:
\begin{lstlisting}[basicstyle=\fontsize{8.5}{9}\ttfamily]
coin : QuantumOp t => IO Bool
coin = do
  [b] <- run (do
           q <- newQubit {t = t}
           q <- applyH q
           r <- measure [q]
           pure r
         )
  pure b
\end{lstlisting}
  The top-level \textbf{do} block simply realises monadic sequencing for the
  standard IO monad.  The \textbf{do} block within the \texttt{run} environment
  is more interesting and crucial for our development.  It performs monadic
  sequencing for the \texttt{QStateT} monad and it represents the simple
  three-step algorithm we just described.  The call to the \texttt{run}
  function executes this algorithm and users obtain the produced classical
  information by storing it in the variable \texttt{b} of type \texttt{Bool}.
  We emphasise that linearity is \emph{enforced} within the \texttt{run} environment
  and this is what ensures the type safety of our approach, e.g. all of the
  following errors are statically detected and rejected by Idris: passing the
  qubit \texttt{q} to a non-linear function, copying the qubit \texttt{q},
  forgetting to measure the qubit \texttt{q} so that it is implicitly
  discarded, etc. For example, if in the above code we replace the last two
  statements in the \texttt{run} environment with "\texttt{pure True}", then
  Idris statically detects this error.
\end{example}

The function \texttt{coin} from Example \ref{ex:coin} is implemented using our
\emph{abstract} interface. This means we can use this function in any
\emph{concrete} implementation of the \texttt{QuantumOp} interface. Since the authors do not have any quantum hardware, we provide one concrete implementation of this interface,
called \texttt{SimulatedOp}, which performs linear-algebraic simulation of all the required operations. Once quantum hardware becomes more readily available,
we can provide additional concrete implementations of the interface. For example, if we wish to use the \texttt{coin} function, then the code:
\begin{lstlisting}
testCoin : IO Bool
testCoin = coin {t = SimulatedOp}
\end{lstlisting} 
defines a new function, called \texttt{testCoin}, which does the same as
\texttt{coin}, but it specifically instructs Idris to use linear-algebraic
simulation. We emphasise that all of our quantum algorithms are written w.r.t.
our abstract interface, so there is no need to reimplement them for any
additional concrete implementations of the interface.

\subsection{Example: Repeat-Until-Success Algorithm}
\label{sub:RUS}

Repeat-until-success (RUS) \cite{RUS} is an algorithm for implementing quantum
unitary operators by using \emph{quantum measurements} and \emph{recursion}. The
main advantage in using RUS over traditional techniques that synthesise unitary
operators, is that RUS usually requires fewer applications of the $T$ gate, which is
expensive in terms of error correction.

\begin{figure}
\begin{lstlisting}[basicstyle=\fontsize{8.5}{9}\ttfamily]
RUS : QuantumOp t => (1 _ : Qubit) ->
      (u' : Unitary 2) -> (e : Unitary 1) ->
      QStateT (t 1) (t 1) Qubit
RUS q u' e = do
  q' <- newQubit
  [q',q] <- applyUnitary [q',q] u'
  b <- measureQubit q'
  if b then do
         [q] <- applyUnitary [q] (adjoint e)
         RUS q u' e
       else pure q 

example_u' : Unitary 2
example_u' = H 0 $ T 0 $ CNOT 0 1 $ H 0 $ CNOT 0 1 $ T 0 $ H 0 IdGate

runRUS : QuantumOp t => IO Bool
runRUS = do
  [b] <- run (do
              q <- newQubit {t = t}
              q <- RUS q example_u' IdGate
              measure [q]
         )
  pure b

testRUS : IO Bool
testRUS = runRUS {t = SimulatedOp}
\end{lstlisting}
\caption{Repeat-until-success algorithm (file: \texttt{RUS.idr}).}
\label{fig:RUS}
\end{figure}

In the simplest case, we wish to realise a fixed
single-qubit unitary operator $U : \mathbb C^2 \to \mathbb C^2.$ The RUS
algorithm is as follows. Given an input qubit $\ket \psi,$ then:
(1) prepare a new qubit in state $\ket 0$;
(2) apply a two-qubit unitary operator $U'$ (chosen in advance depending on $U$);
(3) measure the first qubit;
(4) if the measurement outcome is $0$ (which occurs with probability $p > 0$), then the output state is $U \ket \psi$, as required, and the
algorithm terminates; otherwise the current state is $E \ket \psi$, where $E$
is some other unitary operator (chosen in advance depending on $U$), so we apply $E^\dagger$ to
this state and we go back to step (1).
The unitary operators $U'$ and $E$ are
chosen in advance, depending on $U$, before the algorithm starts so that the above conditions are satisfied.

This process always terminates in state $U \ket \psi$ (provided $p>0$) so RUS
indeed implements the unitary operator $U$.  Note that this is an
\emph{algorithmic} realisation of $U$, not an algebraic one, and so we cannot
write a program of type \texttt{Unitary} that achieves this. Instead, we
represent this as a quantum program in Figure \ref{fig:RUS}. There, \texttt{RUS
q u' e} is the quantum state transformer which \emph{is} the RUS algorithm as
above. The function \texttt{runRUS} simply executes the RUS algorithm on a
qubit in state $\ket 0$, with the unitary operator chosen from \cite[Figure
8]{RUS}, then measures the qubit and returns the outcome. Both of
these functions are written using our abstract interface.  The function
\texttt{testRUS} is the same as \texttt{runRUS}, but it also instructs Idris to use
linear-algebraic simulation for the execution.

\begin{remark}
  It is possible to make the \texttt{RUS} function non-linear in the qubit
  argument and still get a valid function. However, such a function cannot be called
  from the \texttt{run} environment (which enforces linearity) as we do in the
  \texttt{runRUS} function. This leads to an error which is statically detected
  by Idris. Therefore, such a non-linear \texttt{RUS} function would be
  useless, because we can never actually \emph{run} it.
\end{remark}

\section{Variational Quantum Programming}
\label{sec:variational}

In the previous section we saw that \name{} is suitable for writing recursive
and effectful quantum programs that make use of quantum measurements. Moreover,
Idris 2 is an excellent programming language with an advanced type system and
fist-class support for classical programming features. In order to demonstrate
that \name{} is suitable for \emph{variational} quantum programming, we also
have to show that both classical and quantum programming features may be
elegantly combined. This is the purpose of this section and we achieve this
through the ultimate test -- implementing the two most prominent variational
quantum algorithms: the Quantum Algorithm for Approximate Optimisation (QAOA)
\cite{QAOA} and the Variational Quantum Eigensolver (VQE) \cite{vqa1}. To the
best of our knowledge, this is the first implementation of these algorithms
that has been achieved in a type-safe framework.

\subsubsection*{General Framework}

Both variational algorithms presented in this section are trying to
find the minimum (or maximum) eigenvalue of a Hamiltonian.
A Hamiltonian is a Hermitian (i.e., self-adjoint) matrix $H$. Its
minimum eigenvalue is the minimum (real) value $\lambda$ s.t. $H
\ket\psi = \lambda \ket\psi$ for some nonzero vector $\ket\psi$.
As $H$ is unitarily diagonalizable, this is equivalent to the minimum of $\bra\psi H \ket\psi$ for all vectors $\ket\psi$
of norm $1$, where $\bra\psi \eqdef \ket\psi^\dagger$.

Both algorithms proceed in the same way to find this value.  They start with
some assumption on what the vector $\ket\psi$ looks like and usually $\ket \psi$ is
prepared by a quantum circuit that depends on some real parameters $\alpha_1, \ldots,
\alpha_p$. In the VQE algorithm, this is called the \emph{Ansatz}. 

By measuring this state $\ket\psi$, one obtains some information on the value
of $\bra\psi H \ket\psi$. This information can then be fed to a \emph{classical} optimizer to change
the value of the parameters $\alpha_1, \ldots, \alpha_p$ for subsequent execution.

This classical-quantum back and forth is repeated until some satisfactory
termination condition has been satisfied (e.g. simply repeat this $k$ times,
where $k \in \mathbb N$ is some large constant). However, in both algorithms,
there is no guarantee that we will find the minimum eigenvalue.

\subsubsection*{QAOA}

QAOA is a variational algorithm \cite{QAOA} that approximately solves
optimization problems. Let $f: \{0,1\}^n \to \mathbb{R}$ be a function
for which we want to find its minimum. We see $f$ as a diagonal
Hamiltonian over $n$ qubits defined by $H \ket{x} = f(x) \ket{x}$ for
all $x \in \{0,1\}^n$. We are therefore searching for the minimum
eigenvalue of this Hamiltonian.

For QAOA to work, the Hamiltonian $H$ should have a special form so that a
circuit for $e^{\gamma H}$ is easy to make, where $\gamma \in \mathbb R$ is a
parameter (to be tuned) that is used by the algorithm. A well-known and
important example is to compute the maximum cut of an undirected graph, i.e.,
to solve the MAXCUT problem.

Our implementation for QAOA on the MAXCUT problem is presented in the file
\texttt{QAOA.idr} and an excerpt is shown in Figure \ref{fig:qaoa}. The Ansatz depends on the graph $G$ for which  we want the
maximum cut, a depth parameter $p$, and some real parameters  $\beta_i,
\gamma_i$.  The depth parameter $p \in \mathbb N$ is usually fixed to be small,
and we have a guarantee that the results of our algorithm become better
when $p$ becomes larger. The real parameters $\beta_i$ and $\gamma_i$ are used
to determine rotation angles for some of the unitary operators that we
need for constructing the relevant circuits.

Following the general framework, in our implementation, we have a function
\texttt{QAOA\_Unitary}, that takes these parameters as input and produces a
unitary circuit that may be used to prepare the state $\ket\psi$ when applied
to the initial state $\ket 0^{\otimes n}$.
We then measure this state $\ket\psi$ and present the result (a cut of the
graph in the obvious binary encoding) to an Optimiser. Our optimiser is implemented by the function 
\texttt{classicalOptimisation} that uses 
all observable information from all previous runs (which amounts to the values
of the parameters $\beta_i, \gamma_i$ and the value of the cuts that
have been previously obtained through quantum measurements) to compute the subsequent rotation parameters $\beta_i, \gamma_i$ that we will use for the next iteration.
The type of this function indicates that it uses the IO monad: this is because
we wish to allow the function to use probabilistic optimisation algorithm or
even external tools. The simplest implementation of this
function chooses the rotation parameters at random.

The interplay between the classical and the quantum part is presented
in Figure \ref{fig:qaoa}.  The function \texttt{QAOA} takes as input an
integer $k$ representing how many times the whole routine will be
done, the depth $p$ of the circuit, and the graph $G$ on which to
compute the cut. Notice that the call to the quantum processor is isolated inside the \texttt{run} function.

\begin{figure*}
\begin{lstlisting}[basicstyle=\fontsize{8}{9}\ttfamily]
QAOA_Unitary : {n : Nat} -> (betas : Vect p Double) -> (gammas : Vect p Double) -> 
               (graph: Graph n) -> Unitary n

classicalOptimisation : {p : Nat} -> (graph : Graph n) ->
                       (previous_info : Vect k (Vect p Double, Vect p Double, Cut n)) -> 
                       IO (Vect p Double, Vect p Double)

QAOA' : QuantumOp t =>
        {n : Nat} ->
        (k : Nat) -> (p : Nat) -> (graph : Graph n) ->
        IO (Vect k (Vect p Double, Vect p Double, Cut n))
QAOA' 0 p graph = pure []
QAOA' (S k) p graph = do
  previous_info <- QAOA' {t} k p graph 
  (betas, gammas) <- classicalOptimisation graph previous_info
  let circuit = QAOA_Unitary betas gammas graph
  cut <- run (do
              qs <- newQubits {t} n
              qs <- applyUnitary qs circuit 
              measureAll qs
              )
  pure $ (betas, gammas, cut) :: previous_info

QAOA : QuantumOp t => {n : Nat} -> (k : Nat) -> (p : Nat) -> Graph n -> IO (Cut n)
QAOA k p graph = do
  res <- QAOA' {t} k p graph
  let cuts = map (\(_, _, cut) => cut) res
  let (cut,size) = bestCut graph cuts
  pure cut
\end{lstlisting}
\caption{\name{} implementation (excerpt) for the QAOA algorithm solving the MAXCUT problem (file: \texttt{QAOA.idr}).}
\label{fig:qaoa}
\end{figure*}

\subsubsection*{VQE}

In the VQE algorithm \cite{vqa1}, one often assumes that $H$ is a tensor
product of \emph{Pauli matrices} (which we do not define here for simplicity)
and then one can use \emph{Hamiltonian averaging}, i.e., we produce a unitary operator $U_H$
that we apply to $\ket\psi$, with the following property: if we measure the
first bit of $U_H \ket\psi$ and denote by $p_0$ (resp. $p_1$) the probability
of the outcome being 0 (resp. 1), then $\bra\psi H \ket\psi = p_0 - p_1$.  We
follow the same approach for our implementation which is available in the file
\texttt{VQE.idr} (we do not provide an excerpt here for lack of space).

This unitary $U_H$ is built by the function \texttt{encodingUnitary} and the
process to compute $\bra\psi H \ket\psi$ is given in the function
\texttt{computeEnergyPauli}. This function runs
the quantum process \texttt{nSamples} times and returns the difference between
the two possible outcomes (0 and 1).

More generally, any Hamiltonian $H$ can be written as a linear combination of
tensor products of Pauli matrices, $H =\sum_i \alpha_i H_i$,  and one can
compute  $\bra\psi H \ket\psi$ by computing each term $\bra\psi H_i \ket\psi$
independently, see the code for details.

\section{Related Work and Future Work}
\label{sec:related}

In this section we compare \name{} with other existing quantum programming
languages that are implemented in software.  We omit comparisons with quantum
type systems that do not have a software implementation, mostly for brevity,
but also because we do not feel such comparisons are fair towards the type
systems, which are usually much smaller and with fewer features (but also more
formal).

The QWIRE language \cite{qwire,reqwire} is a quantum circuit language that is
embedded in the Coq proof assistant \cite{coq}. In that sense, QWIRE is similar
to \name{}, because it also has access to dependent types (provided by Coq),
and because it also clearly separates the quantum and classical modes of
programming. However, Idris 2 is \emph{not} a proof assistant (it is a
programming language), it does not have access to an interactive proof system
and it is not suitable for formally verifying complex properties of quantum
programs. Indeed, this is what QWIRE excels at. The focus of \name{} is on
\emph{programming} and Idris 2 has better support for classical, quantum and
effectful programming features. In particular, since Coq lacks general
recursion, the RUS algorithm from \secref{sub:RUS} cannot be expressed in QWIRE
and the same is true for many other classical and quantum algorithms that make
use of general recursion or probabilistic effects.

Silq \cite{silq} is a recently described standalone quantum programming language which
is also type-safe (like \name{}) and whose main notable feature is automatic
uncomputation of temporary values. We currently partially support this feature, because we have
clearly identified and separated the reversible fragment of quantum computation
(see the \texttt{Unitary} type) and we can synthesise the required adjoints by
calling the \texttt{adjoint} function. We believe that it is possible to extend
\name{} with full automatic uncomputation as well and we leave this for future work.
Compared to Silq, the main advantage of \name{} is that Idris has much better
support for classical programming features and so we believe that \name{} is a
better choice for \emph{variational} quantum programming, where the classical
part of the algorithm is more complicated and difficult to program. In
addition, Silq does not support general recursion, so it cannot express quantum
algorithms that rely on this (e.g. the RUS algorithm from \secref{sub:RUS}).

The Quipper language \cite{quipper-paper} is a domain specific language (DSL)
embedded in Haskell. \name{} is itself embedded in Idris 2 and clearly Idris 2
has been strongly inspired by Haskell, so the programming styles in \name{} and
Quipper are similar. Quipper has an extensive library of useful quantum
functions and we have not implemented all of them in \name{} yet. We believe
that we can implement most of them and we leave this for future work. The main
advantage that \name{} has over Quipper is that \name{} is type-safe, because
it is implemented in Idris 2, which has dependent types and linearity. Haskell
does not support these features, and because of this, Quipper cannot statically
detect quantum programs that are physically inadmissible, whereas \name{} can
do this, as we already demonstrated in this paper.

Another recent language includes Proto-Quipper-D \cite{proto-quipper-d} which
is a type-safe circuit description language. This language is based on a novel
type system which shows how linearity and dependent types can be combined. A
fundamental difference between Proto-Quipper-D and \name{} is that linearity is
the default mode of operation in Proto-Quipper-D, whereas in \name{} the
default mode is non-linear.
The focus in Proto-Quipper-D is on \emph{circuit}
description and generation and the language currently lacks what is commonly
known as \emph{dynamic lifting}, i.e., the language does not support effectful
quantum measurements and probabilistic effects.  Because of this it cannot be
used for variational quantum programming at present.

Other languages, include Google's Cirq \cite{cirq} (a
set of python libraries), IBM's Qiskit \cite{qiskit} (a set of python
libraries) and Microsoft's Q$\sharp$ \cite{qsharp} (standalone). These
languages offer a wide-range of quantum functions and features, however, none
of them are type-safe and so it is possible to write erroneous quantum programs
which are not detected at compile time. \name{} does not have this problem
and this is indeed its main advantage over them, together with dependent types.


\newpage
\bibliography{refs}


\begin{thebibliography}{28}


\ifx \showCODEN    \undefined \def \showCODEN     #1{\unskip}     \fi
\ifx \showDOI      \undefined \def \showDOI       #1{#1}\fi
\ifx \showISBNx    \undefined \def \showISBNx     #1{\unskip}     \fi
\ifx \showISBNxiii \undefined \def \showISBNxiii  #1{\unskip}     \fi
\ifx \showISSN     \undefined \def \showISSN      #1{\unskip}     \fi
\ifx \showLCCN     \undefined \def \showLCCN      #1{\unskip}     \fi
\ifx \shownote     \undefined \def \shownote      #1{#1}          \fi
\ifx \showarticletitle \undefined \def \showarticletitle #1{#1}   \fi
\ifx \showURL      \undefined \def \showURL       {\relax}        \fi
\providecommand\bibfield[2]{#2}
\providecommand\bibinfo[2]{#2}
\providecommand\natexlab[1]{#1}
\providecommand\showeprint[2][]{arXiv:#2}

\bibitem[\protect\citeauthoryear{Aleksandrowicz, Alexander, Barkoutsos, Bello,
  Ben-Haim, Bucher, Cabrera-Hernández, Carballo-Franquis, Chen, Chen, Chow,
  Córcoles-Gonzales, Cross, Cross, Cruz-Benito, Culver, González, Torre,
  Ding, Dumitrescu, Duran, Eendebak, Everitt, Sertage, Frisch, Fuhrer,
  Gambetta, Gago, Gomez-Mosquera, Greenberg, Hamamura, Havlicek, Hellmers,
  Łukasz Herok, Horii, Hu, Imamichi, Itoko, Javadi-Abhari, Kanazawa, Karazeev,
  Krsulich, Liu, Luh, Maeng, Marques, Martín-Fernández, McClure, McKay,
  Meesala, Mezzacapo, Moll, Rodríguez, Nannicini, Nation, Ollitrault,
  O'Riordan, Paik, Pérez, Phan, Pistoia, Prutyanov, Reuter, Rice, Davila,
  Rudy, Ryu, Sathaye, Schnabel, Schoute, Setia, Shi, Silva, Siraichi,
  Sivarajah, Smolin, Soeken, Takahashi, Tavernelli, Taylor, Taylour, Trabing,
  Treinish, Turner, Vogt-Lee, Vuillot, Wildstrom, Wilson, Winston, Wood, Wood,
  Wörner, Akhalwaya, and Zoufal}{Aleksandrowicz et~al\mbox{.}}{2019}]%
        {qiskit}
\bibfield{author}{\bibinfo{person}{Gadi Aleksandrowicz},
  \bibinfo{person}{Thomas Alexander}, \bibinfo{person}{Panagiotis Barkoutsos},
  \bibinfo{person}{Luciano Bello}, \bibinfo{person}{Yael Ben-Haim},
  \bibinfo{person}{David Bucher}, \bibinfo{person}{Francisco~Jose
  Cabrera-Hernández}, \bibinfo{person}{Jorge Carballo-Franquis},
  \bibinfo{person}{Adrian Chen}, \bibinfo{person}{Chun-Fu Chen},
  \bibinfo{person}{Jerry~M. Chow}, \bibinfo{person}{Antonio~D.
  Córcoles-Gonzales}, \bibinfo{person}{Abigail~J. Cross},
  \bibinfo{person}{Andrew Cross}, \bibinfo{person}{Juan Cruz-Benito},
  \bibinfo{person}{Chris Culver}, \bibinfo{person}{Salvador De La~Puente
  González}, \bibinfo{person}{Enrique De~La Torre}, \bibinfo{person}{Delton
  Ding}, \bibinfo{person}{Eugene Dumitrescu}, \bibinfo{person}{Ivan Duran},
  \bibinfo{person}{Pieter Eendebak}, \bibinfo{person}{Mark Everitt},
  \bibinfo{person}{Ismael~Faro Sertage}, \bibinfo{person}{Albert Frisch},
  \bibinfo{person}{Andreas Fuhrer}, \bibinfo{person}{Jay Gambetta},
  \bibinfo{person}{Borja~Godoy Gago}, \bibinfo{person}{Juan Gomez-Mosquera},
  \bibinfo{person}{Donny Greenberg}, \bibinfo{person}{Ikko Hamamura},
  \bibinfo{person}{Vojtech Havlicek}, \bibinfo{person}{Joe Hellmers},
  \bibinfo{person}{Łukasz Herok}, \bibinfo{person}{Hiroshi Horii},
  \bibinfo{person}{Shaohan Hu}, \bibinfo{person}{Takashi Imamichi},
  \bibinfo{person}{Toshinari Itoko}, \bibinfo{person}{Ali Javadi-Abhari},
  \bibinfo{person}{Naoki Kanazawa}, \bibinfo{person}{Anton Karazeev},
  \bibinfo{person}{Kevin Krsulich}, \bibinfo{person}{Peng Liu},
  \bibinfo{person}{Yang Luh}, \bibinfo{person}{Yunho Maeng},
  \bibinfo{person}{Manoel Marques}, \bibinfo{person}{Francisco~Jose
  Martín-Fernández}, \bibinfo{person}{Douglas~T. McClure},
  \bibinfo{person}{David McKay}, \bibinfo{person}{Srujan Meesala},
  \bibinfo{person}{Antonio Mezzacapo}, \bibinfo{person}{Nikolaj Moll},
  \bibinfo{person}{Diego~Moreda Rodríguez}, \bibinfo{person}{Giacomo
  Nannicini}, \bibinfo{person}{Paul Nation}, \bibinfo{person}{Pauline
  Ollitrault}, \bibinfo{person}{Lee~James O'Riordan}, \bibinfo{person}{Hanhee
  Paik}, \bibinfo{person}{Jesús Pérez}, \bibinfo{person}{Anna Phan},
  \bibinfo{person}{Marco Pistoia}, \bibinfo{person}{Viktor Prutyanov},
  \bibinfo{person}{Max Reuter}, \bibinfo{person}{Julia Rice},
  \bibinfo{person}{Abdón~Rodríguez Davila}, \bibinfo{person}{Raymond
  Harry~Putra Rudy}, \bibinfo{person}{Mingi Ryu}, \bibinfo{person}{Ninad
  Sathaye}, \bibinfo{person}{Chris Schnabel}, \bibinfo{person}{Eddie Schoute},
  \bibinfo{person}{Kanav Setia}, \bibinfo{person}{Yunong Shi},
  \bibinfo{person}{Adenilton Silva}, \bibinfo{person}{Yukio Siraichi},
  \bibinfo{person}{Seyon Sivarajah}, \bibinfo{person}{John~A. Smolin},
  \bibinfo{person}{Mathias Soeken}, \bibinfo{person}{Hitomi Takahashi},
  \bibinfo{person}{Ivano Tavernelli}, \bibinfo{person}{Charles Taylor},
  \bibinfo{person}{Pete Taylour}, \bibinfo{person}{Kenso Trabing},
  \bibinfo{person}{Matthew Treinish}, \bibinfo{person}{Wes Turner},
  \bibinfo{person}{Desiree Vogt-Lee}, \bibinfo{person}{Christophe Vuillot},
  \bibinfo{person}{Jonathan~A. Wildstrom}, \bibinfo{person}{Jessica Wilson},
  \bibinfo{person}{Erick Winston}, \bibinfo{person}{Christopher Wood},
  \bibinfo{person}{Stephen Wood}, \bibinfo{person}{Stefan Wörner},
  \bibinfo{person}{Ismail~Yunus Akhalwaya}, {and} \bibinfo{person}{Christa
  Zoufal}.} \bibinfo{year}{2019}\natexlab{}.
\newblock \bibinfo{booktitle}{\emph{{Qiskit: An Open-source Framework for
  Quantum Computing}}}.
\newblock
\urldef\tempurl%
\url{https://doi.org/10.5281/zenodo.2562111}
\showDOI{\tempurl}


\bibitem[\protect\citeauthoryear{Atkey}{Atkey}{2009}]%
        {indexed-monads}
\bibfield{author}{\bibinfo{person}{Robert Atkey}.}
  \bibinfo{year}{2009}\natexlab{}.
\newblock \showarticletitle{Parameterised notions of computation}.
\newblock \bibinfo{journal}{\emph{J. Funct. Program.}} \bibinfo{volume}{19},
  \bibinfo{number}{3-4} (\bibinfo{year}{2009}), \bibinfo{pages}{335--376}.
\newblock
\urldef\tempurl%
\url{https://doi.org/10.1017/S095679680900728X}
\showDOI{\tempurl}


\bibitem[\protect\citeauthoryear{Atkey}{Atkey}{2018}]%
        {qtt2}
\bibfield{author}{\bibinfo{person}{Robert Atkey}.}
  \bibinfo{year}{2018}\natexlab{}.
\newblock \showarticletitle{Syntax and Semantics of Quantitative Type Theory}.
  In \bibinfo{booktitle}{\emph{Proceedings of the 33rd Annual {ACM/IEEE}
  Symposium on Logic in Computer Science, {LICS} 2018, Oxford, UK, July 09-12,
  2018}}, \bibfield{editor}{\bibinfo{person}{Anuj Dawar} {and}
  \bibinfo{person}{Erich Gr{\"{a}}del}} (Eds.). \bibinfo{publisher}{{ACM}},
  \bibinfo{pages}{56--65}.
\newblock
\urldef\tempurl%
\url{https://doi.org/10.1145/3209108.3209189}
\showDOI{\tempurl}


\bibitem[\protect\citeauthoryear{Benton}{Benton}{1995}]%
        {benton-small}
\bibfield{author}{\bibinfo{person}{P.N. Benton}.}
  \bibinfo{year}{1995}\natexlab{}.
\newblock \showarticletitle{A mixed linear and non-linear logic: Proofs, terms
  and models}. In \bibinfo{booktitle}{\emph{Computer Science Logic: 8th
  Workshop, CSL '94, Selected Papaers}}.
\newblock
\urldef\tempurl%
\url{https://doi.org/10.1007/BFb0022251}
\showDOI{\tempurl}


\bibitem[\protect\citeauthoryear{Benton and Wadler}{Benton and Wadler}{1996}]%
        {benton-wadler}
\bibfield{author}{\bibinfo{person}{P.~N. Benton} {and} \bibinfo{person}{P.
  Wadler}.} \bibinfo{year}{1996}\natexlab{}.
\newblock \showarticletitle{Linear Logic, Monads and the Lambda Calculus}. In
  \bibinfo{booktitle}{\emph{{LICS} 1996}}.
\newblock


\bibitem[\protect\citeauthoryear{Bichsel, Baader, Gehr, and Vechev}{Bichsel
  et~al\mbox{.}}{2020}]%
        {silq}
\bibfield{author}{\bibinfo{person}{Benjamin Bichsel},
  \bibinfo{person}{Maximilian Baader}, \bibinfo{person}{Timon Gehr}, {and}
  \bibinfo{person}{Martin~T. Vechev}.} \bibinfo{year}{2020}\natexlab{}.
\newblock \showarticletitle{Silq: a high-level quantum language with safe
  uncomputation and intuitive semantics}. In
  \bibinfo{booktitle}{\emph{Proceedings of the 41st {ACM} {SIGPLAN}
  International Conference on Programming Language Design and Implementation,
  {PLDI} 2020, London, UK, June 15-20, 2020}},
  \bibfield{editor}{\bibinfo{person}{Alastair~F. Donaldson} {and}
  \bibinfo{person}{Emina Torlak}} (Eds.). \bibinfo{publisher}{{ACM}},
  \bibinfo{pages}{286--300}.
\newblock
\urldef\tempurl%
\url{https://doi.org/10.1145/3385412.3386007}
\showDOI{\tempurl}


\bibitem[\protect\citeauthoryear{Brady}{Brady}{2021}]%
        {idris2}
\bibfield{author}{\bibinfo{person}{Edwin~C. Brady}.}
  \bibinfo{year}{2021}\natexlab{}.
\newblock \showarticletitle{Idris 2: Quantitative Type Theory in Practice}. In
  \bibinfo{booktitle}{\emph{35th European Conference on Object-Oriented
  Programming, {ECOOP} 2021, July 11-17, 2021, Aarhus, Denmark (Virtual
  Conference)}} \emph{(\bibinfo{series}{LIPIcs}, Vol.~\bibinfo{volume}{194})},
  \bibfield{editor}{\bibinfo{person}{Anders M{\o}ller} {and}
  \bibinfo{person}{Manu Sridharan}} (Eds.). \bibinfo{publisher}{Schloss
  Dagstuhl - Leibniz-Zentrum f{\"{u}}r Informatik}, \bibinfo{pages}{9:1--9:26}.
\newblock
\urldef\tempurl%
\url{https://doi.org/10.4230/LIPIcs.ECOOP.2021.9}
\showDOI{\tempurl}


\bibitem[\protect\citeauthoryear{Farhi, Goldstone, and Gutmann}{Farhi
  et~al\mbox{.}}{2014}]%
        {QAOA}
\bibfield{author}{\bibinfo{person}{Edward Farhi}, \bibinfo{person}{Jeffrey
  Goldstone}, {and} \bibinfo{person}{Sam Gutmann}.}
  \bibinfo{year}{2014}\natexlab{}.
\newblock \bibinfo{title}{A Quantum Approximate Optimization Algorithm}.
\newblock
\newblock
\showeprint[arxiv]{1411.4028}~[quant-ph]


\bibitem[\protect\citeauthoryear{Fu, Kishida, Ross, and Selinger}{Fu
  et~al\mbox{.}}{2020}]%
        {proto-quipper-d}
\bibfield{author}{\bibinfo{person}{Peng Fu}, \bibinfo{person}{Kohei Kishida},
  \bibinfo{person}{Neil~J. Ross}, {and} \bibinfo{person}{Peter Selinger}.}
  \bibinfo{year}{2020}\natexlab{}.
\newblock \showarticletitle{A Tutorial Introduction to Quantum Circuit
  Programming in Dependently Typed Proto-Quipper}. In
  \bibinfo{booktitle}{\emph{Reversible Computation - 12th International
  Conference, {RC} 2020, Oslo, Norway, July 9-10, 2020, Proceedings}}
  \emph{(\bibinfo{series}{Lecture Notes in Computer Science},
  Vol.~\bibinfo{volume}{12227})}, \bibfield{editor}{\bibinfo{person}{Ivan
  Lanese} {and} \bibinfo{person}{Mariusz Rawski}} (Eds.).
  \bibinfo{publisher}{Springer}, \bibinfo{pages}{153--168}.
\newblock
\urldef\tempurl%
\url{https://doi.org/10.1007/978-3-030-52482-1\_9}
\showDOI{\tempurl}


\bibitem[\protect\citeauthoryear{Girard}{Girard}{1987}]%
        {girard}
\bibfield{author}{\bibinfo{person}{J.-Y. Girard}.}
  \bibinfo{year}{1987}\natexlab{}.
\newblock \showarticletitle{Linear Logic}.
\newblock \bibinfo{journal}{\emph{Theoretical Computer Science}}
  \bibinfo{volume}{50} (\bibinfo{year}{1987}), \bibinfo{pages}{1 -- 101}.
\newblock


\bibitem[\protect\citeauthoryear{Green, Lumsdaine, Ross, Selinger, and
  Valiron}{Green et~al\mbox{.}}{2013}]%
        {quipper-paper}
\bibfield{author}{\bibinfo{person}{A.~S. Green}, \bibinfo{person}{P.~L.
  Lumsdaine}, \bibinfo{person}{N.~J. Ross}, \bibinfo{person}{P. Selinger},
  {and} \bibinfo{person}{B. Valiron}.} \bibinfo{year}{2013}\natexlab{}.
\newblock \showarticletitle{Quipper: a scalable quantum programming language}.
  In \bibinfo{booktitle}{\emph{{PLDI}}}. \bibinfo{publisher}{{ACM}},
  \bibinfo{pages}{333--342}.
\newblock


\bibitem[\protect\citeauthoryear{Jia, Kornell, Lindenhovius, Mislove, and
  Zamdzhiev}{Jia et~al\mbox{.}}{2022}]%
        {vqpl}
\bibfield{author}{\bibinfo{person}{Xiaodong Jia}, \bibinfo{person}{Andre
  Kornell}, \bibinfo{person}{Bert Lindenhovius}, \bibinfo{person}{Michael~W.
  Mislove}, {and} \bibinfo{person}{Vladimir Zamdzhiev}.}
  \bibinfo{year}{2022}\natexlab{}.
\newblock \showarticletitle{Semantics for Variational Quantum Programming}.
\newblock \bibinfo{journal}{\emph{Proc. {ACM} Program. Lang.}}
  \bibinfo{volume}{6}, \bibinfo{number}{{POPL}} (\bibinfo{year}{2022}).
\newblock
\showeprint[arxiv]{2107.13347}
\newblock
\shownote{To appear.}


\bibitem[\protect\citeauthoryear{Mac~Lane}{Mac~Lane}{1998}]%
        {maclane}
\bibfield{author}{\bibinfo{person}{Saunders Mac~Lane}.}
  \bibinfo{year}{1998}\natexlab{}.
\newblock \bibinfo{booktitle}{\emph{{Categories for the Working Mathematician
  (2nd ed.)}}}.
\newblock \bibinfo{publisher}{Springer}.
\newblock


\bibitem[\protect\citeauthoryear{McBride}{McBride}{2016}]%
        {qtt1}
\bibfield{author}{\bibinfo{person}{Conor McBride}.}
  \bibinfo{year}{2016}\natexlab{}.
\newblock \showarticletitle{I Got Plenty o' Nuttin'}. In
  \bibinfo{booktitle}{\emph{A List of Successes That Can Change the World -
  Essays Dedicated to Philip Wadler on the Occasion of His 60th Birthday}}
  \emph{(\bibinfo{series}{Lecture Notes in Computer Science},
  Vol.~\bibinfo{volume}{9600})}, \bibfield{editor}{\bibinfo{person}{Sam
  Lindley}, \bibinfo{person}{Conor McBride}, \bibinfo{person}{Philip~W.
  Trinder}, {and} \bibinfo{person}{Donald Sannella}} (Eds.).
  \bibinfo{publisher}{Springer}, \bibinfo{pages}{207--233}.
\newblock
\urldef\tempurl%
\url{https://doi.org/10.1007/978-3-319-30936-1\_12}
\showDOI{\tempurl}


\bibitem[\protect\citeauthoryear{McClean, Romero, Babbush, and
  Aspuru-Guzik}{McClean et~al\mbox{.}}{2016}]%
        {vqa2}
\bibfield{author}{\bibinfo{person}{Jarrod~R McClean}, \bibinfo{person}{Jonathan
  Romero}, \bibinfo{person}{Ryan Babbush}, {and} \bibinfo{person}{Al{\'a}n
  Aspuru-Guzik}.} \bibinfo{year}{2016}\natexlab{}.
\newblock \showarticletitle{The theory of variational hybrid quantum-classical
  algorithms}.
\newblock \bibinfo{journal}{\emph{New Journal of Physics}}
  \bibinfo{volume}{18}, \bibinfo{number}{2} (\bibinfo{year}{2016}),
  \bibinfo{pages}{023023}.
\newblock


\bibitem[\protect\citeauthoryear{Nielsen and Chuang}{Nielsen and
  Chuang}{2010}]%
        {nielsen-chuang}
\bibfield{author}{\bibinfo{person}{Michael~A. Nielsen} {and}
  \bibinfo{person}{Isaac~L. Chuang}.} \bibinfo{year}{2010}\natexlab{}.
\newblock \bibinfo{booktitle}{\emph{Quantum Computation and Quantum
  Information: 10th Anniversary Edition}}.
\newblock \bibinfo{publisher}{Cambridge University Press}.
\newblock
\urldef\tempurl%
\url{https://doi.org/10.1017/CBO9780511976667}
\showDOI{\tempurl}


\bibitem[\protect\citeauthoryear{Paetznick and Svore}{Paetznick and
  Svore}{2014}]%
        {RUS}
\bibfield{author}{\bibinfo{person}{Adam Paetznick} {and}
  \bibinfo{person}{Krysta~M. Svore}.} \bibinfo{year}{2014}\natexlab{}.
\newblock \showarticletitle{Repeat-until-Success: Non-Deterministic
  Decomposition of Single-Qubit Unitaries}.
\newblock \bibinfo{journal}{\emph{Quantum Info. Comput.}} \bibinfo{volume}{14},
  \bibinfo{number}{15–16} (\bibinfo{date}{Nov.} \bibinfo{year}{2014}),
  \bibinfo{pages}{1277–1301}.
\newblock
\showISSN{1533-7146}


\bibitem[\protect\citeauthoryear{Paykin, Rand, and Zdancewic}{Paykin
  et~al\mbox{.}}{2017}]%
        {qwire}
\bibfield{author}{\bibinfo{person}{J. Paykin}, \bibinfo{person}{R. Rand}, {and}
  \bibinfo{person}{S. Zdancewic}.} \bibinfo{year}{2017}\natexlab{}.
\newblock \showarticletitle{{QWIRE:} a core language for quantum circuits}. In
  \bibinfo{booktitle}{\emph{{POPL}}}. \bibinfo{publisher}{{ACM}},
  \bibinfo{pages}{846--858}.
\newblock


\bibitem[\protect\citeauthoryear{P{\'{e}}choux, Perdrix, Rennela, and
  Zamdzhiev}{P{\'{e}}choux et~al\mbox{.}}{2020}]%
        {qpl-fossacs}
\bibfield{author}{\bibinfo{person}{Romain P{\'{e}}choux},
  \bibinfo{person}{Simon Perdrix}, \bibinfo{person}{Mathys Rennela}, {and}
  \bibinfo{person}{Vladimir Zamdzhiev}.} \bibinfo{year}{2020}\natexlab{}.
\newblock \showarticletitle{Quantum Programming with Inductive Datatypes:
  Causality and Affine Type Theory}. In \bibinfo{booktitle}{\emph{Foundations
  of Software Science and Computation Structures - 23rd International
  Conference, {FOSSACS} 2020}} \emph{(\bibinfo{series}{Lecture Notes in
  Computer Science}, Vol.~\bibinfo{volume}{12077})}.
  \bibinfo{publisher}{Springer}, \bibinfo{pages}{562--581}.
\newblock
\urldef\tempurl%
\url{https://doi.org/10.1007/978-3-030-45231-5\_29}
\showDOI{\tempurl}


\bibitem[\protect\citeauthoryear{Peruzzo, McClean, Shadbolt, Yung, Zhou, Love,
  Aspuru-Guzik, and O’brien}{Peruzzo et~al\mbox{.}}{2014}]%
        {vqa1}
\bibfield{author}{\bibinfo{person}{Alberto Peruzzo}, \bibinfo{person}{Jarrod
  McClean}, \bibinfo{person}{Peter Shadbolt}, \bibinfo{person}{Man-Hong Yung},
  \bibinfo{person}{Xiao-Qi Zhou}, \bibinfo{person}{Peter~J Love},
  \bibinfo{person}{Al{\'a}n Aspuru-Guzik}, {and} \bibinfo{person}{Jeremy~L
  O’brien}.} \bibinfo{year}{2014}\natexlab{}.
\newblock \showarticletitle{A variational eigenvalue solver on a photonic
  quantum processor}.
\newblock \bibinfo{journal}{\emph{Nature communications}} \bibinfo{volume}{5},
  \bibinfo{number}{1} (\bibinfo{year}{2014}), \bibinfo{pages}{1--7}.
\newblock


\bibitem[\protect\citeauthoryear{Rand, Paykin, Lee, and Zdancewic}{Rand
  et~al\mbox{.}}{2018}]%
        {reqwire}
\bibfield{author}{\bibinfo{person}{Robert Rand}, \bibinfo{person}{Jennifer
  Paykin}, \bibinfo{person}{Dong{-}Ho Lee}, {and} \bibinfo{person}{Steve
  Zdancewic}.} \bibinfo{year}{2018}\natexlab{}.
\newblock \showarticletitle{ReQWIRE: Reasoning about Reversible Quantum
  Circuits}. In \bibinfo{booktitle}{\emph{Proceedings 15th International
  Conference on Quantum Physics and Logic, {QPL} 2018, Halifax, Canada, 3-7th
  June 2018}} \emph{(\bibinfo{series}{{EPTCS}}, Vol.~\bibinfo{volume}{287})},
  \bibfield{editor}{\bibinfo{person}{Peter Selinger} {and}
  \bibinfo{person}{Giulio Chiribella}} (Eds.). \bibinfo{pages}{299--312}.
\newblock
\urldef\tempurl%
\url{https://doi.org/10.4204/EPTCS.287.17}
\showDOI{\tempurl}


\bibitem[\protect\citeauthoryear{Selinger and Valiron}{Selinger and
  Valiron}{2006}]%
        {quantum-lambda}
\bibfield{author}{\bibinfo{person}{P. Selinger} {and} \bibinfo{person}{B.
  Valiron}.} \bibinfo{year}{2006}\natexlab{}.
\newblock \showarticletitle{A lambda calculus for quantum computation with
  classical control}.
\newblock \bibinfo{journal}{\emph{Mathematical Structures in Computer Science}}
  \bibinfo{volume}{16}, \bibinfo{number}{3} (\bibinfo{year}{2006}),
  \bibinfo{pages}{527--552}.
\newblock


\bibitem[\protect\citeauthoryear{Seo}{Seo}{2017}]%
        {indexed-monads-haskell}
\bibfield{author}{\bibinfo{person}{Kwang~Yul Seo}.}
  \bibinfo{year}{2017}\natexlab{}.
\newblock \bibinfo{title}{Indexed State Monad Blog Post}.
\newblock
  \bibinfo{howpublished}{\url{https://kseo.github.io/posts/2017-01-12-indexed-monads.html}}.
\newblock
\newblock
\shownote{Accessed: 13.08.2021.}


\bibitem[\protect\citeauthoryear{Shor}{Shor}{1999}]%
        {shor}
\bibfield{author}{\bibinfo{person}{Peter~W. Shor}.}
  \bibinfo{year}{1999}\natexlab{}.
\newblock \showarticletitle{{Polynomial-Time Algorithms for Prime Factorization
  and Discrete Logarithms on a Quantum Computer}}.
\newblock \bibinfo{journal}{\emph{SIAM Rev.}} \bibinfo{volume}{41},
  \bibinfo{number}{2} (\bibinfo{year}{1999}), \bibinfo{pages}{303--332}.
\newblock
\urldef\tempurl%
\url{https://doi.org/10.1137/S0036144598347011}
\showDOI{\tempurl}


\bibitem[\protect\citeauthoryear{Svore, Geller, Troyer, Azariah, Granade, Heim,
  Kliuchnikov, Mykhailova, Paz, and Roetteler}{Svore et~al\mbox{.}}{2018}]%
        {qsharp}
\bibfield{author}{\bibinfo{person}{Krysta Svore}, \bibinfo{person}{Alan
  Geller}, \bibinfo{person}{Matthias Troyer}, \bibinfo{person}{John Azariah},
  \bibinfo{person}{Christopher Granade}, \bibinfo{person}{Bettina Heim},
  \bibinfo{person}{Vadym Kliuchnikov}, \bibinfo{person}{Mariia Mykhailova},
  \bibinfo{person}{Andres Paz}, {and} \bibinfo{person}{Martin Roetteler}.}
  \bibinfo{year}{2018}\natexlab{}.
\newblock \showarticletitle{Q\#: Enabling Scalable Quantum Computing and
  Development with a High-Level DSL}. In \bibinfo{booktitle}{\emph{Proceedings
  of the Real World Domain Specific Languages Workshop 2018}} (Vienna, Austria)
  \emph{(\bibinfo{series}{RWDSL2018})}. \bibinfo{publisher}{Association for
  Computing Machinery}, \bibinfo{address}{New York, NY, USA}, Article
  \bibinfo{articleno}{7}, \bibinfo{numpages}{10}~pages.
\newblock
\showISBNx{9781450363556}
\urldef\tempurl%
\url{https://doi.org/10.1145/3183895.3183901}
\showDOI{\tempurl}


\bibitem[\protect\citeauthoryear{Team}{Team}{2021a}]%
        {coq}
\bibfield{author}{\bibinfo{person}{Coq~Development Team}.}
  \bibinfo{year}{2021}\natexlab{a}.
\newblock \bibinfo{title}{The Coq Proof Assistant Reference Manual}.
\newblock
  \bibinfo{howpublished}{\url{https://coq.inria.fr/distrib/current/refman/}}.
\newblock
\newblock
\shownote{Accessed: 19.11.2021.}


\bibitem[\protect\citeauthoryear{Team}{Team}{2021b}]%
        {cirq}
\bibfield{author}{\bibinfo{person}{Google AI~Quantum Team}.}
  \bibinfo{year}{2021}\natexlab{b}.
\newblock \bibinfo{title}{Cirq}.
\newblock \bibinfo{howpublished}{\url{https://quantumai.google/cirq}}.
\newblock
\newblock
\shownote{Accessed: 13.08.2021.}


\bibitem[\protect\citeauthoryear{Wootters and Zurek}{Wootters and
  Zurek}{1982}]%
        {no-cloning}
\bibfield{author}{\bibinfo{person}{William~K Wootters} {and}
  \bibinfo{person}{Wojciech~H Zurek}.} \bibinfo{year}{1982}\natexlab{}.
\newblock \showarticletitle{A single quantum cannot be cloned}.
\newblock \bibinfo{journal}{\emph{Nature}} \bibinfo{volume}{299},
  \bibinfo{number}{5886} (\bibinfo{year}{1982}), \bibinfo{pages}{802--803}.
\newblock


\end{thebibliography}

\end{document}